\documentclass[journal]{new-aiaa}

\usepackage[utf8]{inputenc}
\usepackage{textcomp}
\usepackage{graphicx}
\usepackage{amsmath}

\usepackage{amsthm}
\usepackage{amsfonts}
\usepackage{mathtools}
\usepackage{bm}
\usepackage[version=4]{mhchem}
\usepackage{siunitx}
\usepackage{longtable,tabularx}
\usepackage{booktabs}
\usepackage{caption}
\usepackage{subcaption}
\DeclareCaptionLabelFormat{noparens}{#2)}
\usepackage{algorithm}
\usepackage{algpseudocode}
\usepackage{hyperref}
\newcommand{\R}{\mathbb{R}}
\newcommand{\Sph}{\mathbb{S}} 
\newcommand{\SO}{\mathrm{SO}}
\newcommand{\so}{\mathfrak{so}}

\newcommand{\vct}[1]{\mathbf{#1}}
\newcommand{\mat}[1]{\mathbf{#1}}
\newcommand{\T}{\mathsf{T}}

\DeclareMathOperator{\trace}{tr}

\DeclareMathOperator{\atanTwo}{atan2}

\newcommand{\skewmat}[1]{\left[#1\right]_\times}

\newcommand{\Umax}{U_{\max}}
\newcommand{\ug}{u_g}

\newcommand{\Rturn}{r} 

\newcommand{\ethree}{\vct{e}_3}

\newtheorem{theorem}{Theorem}
\newtheorem{lemma}[theorem]{Lemma}

\theoremstyle{definition}

\theoremstyle{remark}

\title{An Analytic Solution to the Optimal Spherical Dubins Path Problem with Geodesic Curvature Constraints}

\author{Linhong Li\footnote{Ph.D. Student, College of Aerospace Science and Engineering.}
and Qi Feng\footnote{Ph.D. Student, College of Aerospace Science and Engineering.}}
\affil{National University of Defense Technology, Changsha, Hunan, 410073, People's Republic of China}
\affil{State Key Laboratory of Space System Operation and Control, Changsha, Hunan, 410073, People's Republic of China}
\author{Yangang Liang\footnote{Professor, College of Aerospace Science and Engineering, \texttt{Liangyg@nudt.edu.cn} (Corresponding Author).}}
\affil{National University of Defense Technology, Changsha, Hunan, 410073, People's Republic of China}
\affil{State Key Laboratory of Space System Operation and Control, Changsha, Hunan, 410073, People's Republic of China}
\author{Kebo Li\footnote{Associate Professor, College of Aerospace Science and Engineering.}}
\affil{National University of Defense Technology, Changsha, Hunan, 410073, People's Republic of China}
\affil{State Key Laboratory of Space System Operation and Control, Changsha, Hunan, 410073, People's Republic of China}

\begin{document}
\maketitle

\begin{abstract}
Computing shortest paths for curvature-constrained Dubins vehicles on the unit sphere is fundamental to many engineering applications, including long-range flight planning, persistent surveillance patterns, and global routing problems where great circles are natural routes. Numerical optimization methods on $\SO(3)$ suffer from sensitivity to initialization, may converge to local minima, and often miss feasible solution branches. This paper proposes a unified analytic computational approach for spherical Dubins CGC and CCC paths that overcomes these limitations. By exploiting the axis-fixing property of rotations and developing a closed-form back-substitution method using geometric projection, the three-dimensional boundary value problem is reduced to solving a quadratic polynomial equation. The proposed analytic solver achieves machine precision accuracy with errors on the order of $10^{-16}$, is approximately $717$ times faster than numerical methods under the same computational environment, and systematically enumerates all feasible solution branches without requiring exhaustive multi-start initialization. The method provides closed-form solutions for optimal path computation in the regime where turning radius $\Rturn \in (0, 1/2]$, corresponding to $U_{\max} \geq \sqrt{3}$.
\end{abstract}

\section*{Nomenclature}
{\renewcommand\arraystretch{1.0}
\noindent\begin{longtable*}{@{}l @{\quad=\quad} l@{}}
$\Sph^2$ & unit sphere in three-dimensional Euclidean space \\
$\SO(3)$ & special orthogonal group, rotation matrices \\
$\so(3)$ & Lie algebra of $\SO(3)$, skew symmetric matrices \\
$\vct{X}(s)$ & position on the unit sphere \\
$\vct{T}(s)$ & unit tangent vector, velocity direction \\
$\vct{N}(s)$ & normal vector completing Sabban frame \\
$g(s)$ & moving frame in $\SO(3)$ \\
$R_0$ & initial frame in $\SO(3)$ \\
$R_f$ & desired terminal frame in $\SO(3)$ \\
$R$ & normalized target rotation \\
$\Omega(\ug)$ & generator in $\so(3)$ for the left invariant system \\
$\ug(s)$ & geodesic curvature control input \\
$H$ & Hamiltonian for the optimal control problem \\
$\lambda_1(s),\lambda_2(s),\lambda_3(s)$ & costates associated with position, tangent, and normal vectors \\
$A(s)$ & switching function for the control \\
$B(s),C(s)$ & auxiliary costate scalars \\
$\vct{r}(\vct{s})$ & residual vector in $\R^3$ \\
$J(\vct{s})$ & Jacobian of residual vector with respect to segment lengths \\
$\vct{e}_j$ & $j$th standard basis vector in $\R^3$ \\
$h_j$ & finite difference step size for the $j$th component \\
$\bm{\delta}$ & least squares update step for the segment lengths \\
$\Umax$ & geodesic curvature bound \\
$\Rturn$ & tight turn radius on the unit sphere \\
$L_{\mathrm{tot}}$ & total path length \\
$L,R$ & left and right tight turns \\
\end{longtable*}}

\section{Introduction}
\lettrine{T}{he} problem of planning shortest paths for curvature constrained vehicles has a long history in guidance, robotics, and optimal control theory.
In the classical planar setting, Dubins established that the shortest forward-only path between two configurations $(x, y, \theta)$ for a vehicle with minimum turning radius $r$ consists of at most three segments---either circular arcs at minimum turning radius or straight line segments \cite{dubins1957curves}.
Using shorthand notation where $C$ denotes a circular arc and $S$ denotes a straight segment, optimal paths are of form $CSC$ or $CCC$.
Reeds and Shepp extended this to allow reverse motion, yielding a different but closely related family of paths \cite{reeds1990optimal}.
Subsequent work refined these results using Pontryagin's Maximum Principle \cite{boissonat} and geometric control techniques \cite{sussman_geometric_examples}, while variants including weighted objectives \cite{weighted_Markov_Dubins}, asymmetric turns \cite{sinistral/dextral}, and target interception \cite{dubins_circle} have been studied.

Many engineering problems are inherently three-dimensional and surface-constrained, requiring extensions beyond the planar framework.
Examples include long-range flight over the Earth, persistent surveillance patterns, and global routing where great circles are natural routes.
In such settings, geodesic curvature on the underlying manifold generalizes planar curvature.
Sussmann considered 3D paths with curvature constraints, showing optimal trajectories are helicoidal arcs or concatenations of at most three segments \cite{Sussmann1995CDC}, while Chitour and Sigalotti provided existence conditions for surfaces of non-negative curvature \cite{Chitour2005DubinsPO}.

For motion constrained to a sphere, a natural generalization uses geodesic curvature, where zero geodesic curvature corresponds to great circular arcs ($G$).
Monroy-Pérez showed planar Dubins results extend to a unit sphere for the specific case $r = 1/\sqrt{2}$ \cite{monroy}.
This work established that in regimes $0 < r \leq 1/2$, shortest paths are concatenations of at most three arcs, with candidate types $CCC$, $CGC$, $CC$, $CG$, $GC$, $C$, or $G$ \cite{darbha2023optimal}.
The Sabban frame represents the vehicle configuration as a rotation matrix in $\SO(3)$, enabling formulation as an optimal control problem on a Lie group.
Kumar et al.\ showed equivalence between moving frame and coordinate-based descriptions \cite{kumar2024equivalence}, while time-optimal control on $\SO(3)$ \cite{time_optimal_synthesis_SO3,time_optimal_control_satellite} and extensions with free terminal heading \cite{free_terminal_sphere} have been investigated.

A practical challenge is that computing all feasible candidates and selecting the shortest must be reliable and predictable in runtime.
Direct numerical shooting on $\SO(3)$ can be sensitive to initialization and may converge to local minima.
For 3D CSC paths, numerical methods have been developed \cite{hota2010CDC,hota2010optimal}, with approaches splitting the problem into 2D paths plus altitude \cite{vana2020,yuwang2015} or using exhaustive search \cite{Wang2021,Wang2022}, but these are resolution-limited or slow.
For spherical problems with wind, numerical techniques exist \cite{Bakolas} but lack geodesic curvature constraints in the same framework.

Analytic approaches offer advantages: they provide all solutions, have consistent runtime, and achieve machine-precision accuracy.
One direction views boundary conditions as inverse kinematics constraints for serial mechanisms.
Recent work on 3D CSC paths demonstrates this approach \cite{baez2024dubins3d}, encoding Dubins segments as motions of an RRPRR manipulator to reduce boundary conditions to scalar constraints with closed-form back substitution, finding up to seven solutions with orders-of-magnitude speedup.

In this paper, we study the spherical Dubins problem in the Sabban frame formulation, representing configuration as $g(s) \in \SO(3)$.
We develop a unified analytic method leveraging Lie group structure and matrix exponentials for constant control segments, focusing on practical computation of both CGC and CCC candidates through unified one-dimensional reduction with closed-form reconstruction.
The main contributions of this paper lie in two folds:

1) The three-dimensional Dubins path boundary value problem is formulated from a Lie group perspective, representing the vehicle configuration as a rotation matrix in $\SO(3)$ and leveraging the Sabban frame dynamics to express the boundary condition as an exponential product equation on the Lie group.

2) Based on the geometric properties of rotation matrices, the complex three-dimensional boundary value problem is reduced to solving a simple quadratic polynomial equation. This reduction exploits the axis-fixing property of rotations and dot product invariance to eliminate two of the three unknown segment parameters, converting the boundary constraint into a scalar trigonometric equation that admits a closed-form solution through half-angle substitution.

The structure of the remaining sections is as follows: Section II formulates the problem and presents the Sabban frame model on $\SO(3)$; Section III reviews a numerical least squares approach for comparison; Section IV develops the unified analytic solution method with elimination strategy and back-substitution procedure; Section V presents numerical experiments validating the method; and finally, Section VI presents the conclusions.

\section{Problem Statement and Models}
In this section, we first present the Sabban frame model for motion on $\Sph^2$ and its compact representation on $\SO(3)$.
We then formulate the corresponding shortest path boundary value problem with a bound on geodesic curvature.
\subsection{Sabban frame model on $\Sph^2$}
Many 3D guidance and planning problems are intrinsically surface constrained even when the vehicle evolves in three dimensional space.
Examples include long range fixed wing flight over the Earth, which is often approximated locally by a sphere, persistent loitering and surveillance patterns over large geographic areas, and global routing problems where the shortest unconstrained routes are great circle segments.
In these settings, a bound on curvature or turn rate arises naturally from lateral acceleration limits, bank angle limits, and actuator constraints.
The planning objective is then to connect two configurations defined by position and heading using a shortest or time minimal trajectory that respects this bound.
While many practical pipelines approximate the surface locally by a tangent plane and use planar Dubins primitives, this approximation can break down over long distances, near the poles, or whenever global feasibility or optimality depends on curvature of the underlying manifold.
This motivates posing a spherical Dubins problem directly on $\Sph^2$, where the relevant curvature notion is geodesic curvature, which measures how much a curve deviates from the surface geodesics, namely great circles.

\begin{figure}[htbp]
\centering
\includegraphics[width=0.6\textwidth]{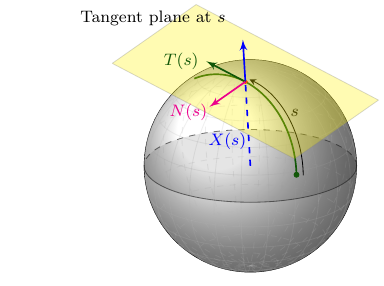}
\caption{Illustration of the Sabban frame vectors $\vct{X}(s)$, $\vct{T}(s)$, and $\vct{N}(s)$ on the unit sphere.
The position vector $\vct{X}(s)$ lies on the sphere, the tangent vector $\vct{T}(s)$ is in the tangent plane, and the normal vector $\vct{N}(s)=\vct{X}(s)\times\vct{T}(s)$ completes the orthonormal frame\cite{darbha2023optimal}.}
\label{fig:sabban_frame}
\end{figure}

From a modeling standpoint, differential geometry provides a natural language for such problems.
At each point on $\Sph^2$, the vehicle's instantaneous direction lies in the tangent plane.
The surface normal and tangent vectors evolve along the path, and curvature constraints can be expressed intrinsically and independent of coordinates.
A convenient way to track both position and heading is to attach a moving orthonormal frame to the vehicle.
As shown in Fig.~\ref{fig:sabban_frame}, on the sphere, the classical \emph{Sabban frame} provides exactly this: the position vector $\vct{X}(s)\in\Sph^2$, the unit tangent $\vct{T}(s)$, and the normal completing the frame $\vct{N}(s)=\vct{X}(s)\times\vct{T}(s)$.

Collecting these vectors as columns yields a rotation matrix $g(s)\in\SO(3)$, i.e., an element of a Lie group.
This representation turns the geometric evolution of the vehicle into a left invariant system on $\SO(3)$.
As a result, tools from Lie groups, including the exponential map and axis angle representations, can be leveraged for both analysis and computation.

We therefore consider unit speed motion on $\Sph^2$ parameterized by arc length $s$.
Let $\vct{X}(s)\in\Sph^2$ denote position, $\vct{T}(s)=d\vct{X}/ds$ the unit tangent, and $\vct{N}(s)=\vct{X}(s)\times\vct{T}(s)$.
With geodesic curvature control $\ug(s)$ bounded by $|\ug(s)|\le \Umax$, the Sabban frame dynamics are
\begin{equation}
\label{eq:sabban}
\frac{d\vct{X}}{ds}=\vct{T},\qquad
\frac{d\vct{T}}{ds}=-\vct{X}+\ug\,\vct{N},\qquad
\frac{d\vct{N}}{ds}=-\ug\,\vct{T}.
\end{equation}
Define the moving frame $g(s)=[\vct{X}(s),\vct{T}(s),\vct{N}(s)]\in\SO(3)$.
Then the dynamics can be written as a left invariant system on $\SO(3)$:
\begin{equation}
\label{eq:lie}
\frac{dg}{ds}=g\,\Omega(\ug),
\end{equation}
where $\Omega(\ug)\in\so(3)$ is the generator matrix.
For the Sabban frame dynamics in \eqref{eq:sabban}, the generator $\Omega(\ug)$ can be written explicitly as the skew symmetric matrix
\begin{equation}
\label{eq:Omega_matrix}
\Omega(\ug)=
\begin{bmatrix}
0 & -1 & 0 \\
1 & 0 & -\ug \\
0 & \ug & 0
\end{bmatrix}
\in\so(3).
\end{equation}

The term left invariant refers to an invariance with respect to left multiplication on the group.
Concretely, if $g(s)$ satisfies \eqref{eq:lie} for a given control $\ug(\cdot)$, then for any fixed rotation $Q\in\SO(3)$ the trajectory $Qg(s)$ satisfies the same equation with the same control.
Equivalently, the velocity field at state $g$ is obtained by multiplying a generator $\Omega(\ug)$ that does not depend on $g$.
This viewpoint is useful because on any interval where $\ug$ is constant, the solution has the closed form
\begin{equation}
\label{eq:g_closed_form}
g(s)=g(0)\exp\!\left(s\,\Omega(\ug)\right).
\end{equation}

\subsection{Spherical Dubins boundary value problem}
The planning task is posed in the same spirit as the classical planar Dubins problem \cite{dubins1957curves}, but with the geometry of $\Sph^2$ built in.
Given an initial configuration and a desired terminal configuration, we seek the shortest admissible path subject to a bound on geodesic curvature.
This bound captures limited maneuverability and leads to a small, structured candidate set of optimal segments.

To connect these conclusions to the present notation, we briefly outline the Pontryagin argument that yields the candidate control values.
Introduce costates $\lambda_1(s),\lambda_2(s),\lambda_3(s)\in\R^3$ associated with the state equations for $\vct{X},\vct{T},\vct{N}$ in \eqref{eq:sabban}.
The Hamiltonian can be written as
\begin{equation}
\label{eq:H_ABC}
H = 1 + \langle \lambda_1,\vct{T}\rangle + \langle \lambda_2,-\vct{X}+\ug\,\vct{N}\rangle + \langle \lambda_3,-\ug\,\vct{T}\rangle
= 1 + C + \ug\,A,
\end{equation}
where we define the scalar functions
Let $A$, $B$, and $C$ denote the scalar functions
\begin{equation}
\label{eq:ABC_def}
A = \langle \lambda_2,\vct{N}\rangle-\langle \lambda_3,\vct{T}\rangle,\qquad
B = \langle \lambda_3,\vct{X}\rangle-\langle \lambda_1,\vct{N}\rangle,\qquad
C = \langle \lambda_1,\vct{T}\rangle-\langle \lambda_2,\vct{X}\rangle.
\end{equation}
Along an extremal these satisfy the adjoint relations
\begin{equation}
\label{eq:ABC_ode}
\frac{dA}{ds}=B,\qquad
\frac{dB}{ds}=-A+\ug\,C,\qquad
\frac{dC}{ds}=-\ug\,B.
\end{equation}
Since the Hamiltonian is affine in $\ug$ and the admissible set is the interval $[-\Umax,\Umax]$, minimizing $H=1+C+\ug A$ yields the saturation law
\begin{equation}
\label{eq:ug_from_A}
\ug(s)=
\begin{cases}
-\Umax, & A(s)>0,\\
\ \Umax, & A(s)<0,
\end{cases}
\end{equation}
with $\ug$ undetermined when $A(s)=0$.
However, if $A(s)\equiv 0$ on an interval, then $dA/ds\equiv 0$ implies $B(s)\equiv 0$ and $dC/ds\equiv 0$.
Using $H\equiv 0$ on an extremal gives $C\equiv -1$ on that interval, and then $dB/ds\equiv 0$ implies $-A+\ug C\equiv 0$ so $\ug\equiv 0$.
Consequently, the optimal control takes values in the set $\{-\Umax,0,\Umax\}$.

These constant control values correspond to the candidate geometric segments.
If $\ug(s)\equiv 0$ on an interval, then $d\vct{N}/ds=0$ and $\vct{N}$ is constant, so $\vct{X}(s)$ lies in a fixed plane and traces an arc of a great circle.
If $\ug(s)\equiv U$ is constant with $U\neq 0$, differentiating the $\vct{T}$ equation in \eqref{eq:sabban} and using \eqref{eq:sabban} yields
\begin{equation}
\label{eq:T_second_order}
\frac{d^2 \vct{T}}{ds^2} + \left(1+U^2\right)\vct{T}=0.
\end{equation}
Hence $\vct{T}$ evolves periodically with angular frequency $\sqrt{1+U^2}$, and the corresponding curve is a small circle arc of radius $r=1/\sqrt{1+U^2}$.
In particular, saturation at $U=\pm U_{\max}$ yields the tight turn radius
\begin{equation}
\label{eq:r_from_umax}
\Rturn=\frac{1}{\sqrt{1+U_{\max}^2}}.
\end{equation}
Thus, an optimal trajectory is a concatenation of great circle arcs and tight small circle arcs, which motivates the $G$ and $C$ building blocks and the candidate path types used later.

The following lemma characterizes the structure of optimal paths in the regime of interest.
\begin{lemma}
\label{lem:optimal_path_types}
If $0 < \Rturn \le 1/2$, the optimal path can only be of one of the following types: $CGC$, $CCC$, $CG$, $GC$, $CC$, $C$, or $G$, where $C$ denotes a tight small circle arc of radius $\Rturn$ and $G$ denotes a great circle arc.\cite{darbha2023optimal}
\end{lemma}

On the computational side, formulating the boundary value problem directly on $\SO(3)$ avoids coordinate singularities and allows us to express segment motions via matrix exponentials.
Each constant control segment corresponds to an exponential $\exp(s\,\Omega(u))$, which is a finite rotation about a fixed axis, and concatenations become products of such exponentials.
This is precisely the algebraic structure we exploit to reduce the CGC solve to one scalar equation with closed form back substitution.

Given an initial frame $g(0)=R_0\in\SO(3)$ and a desired final frame $g(L_{\mathrm{tot}})=R_f\in\SO(3)$, the spherical Dubins problem can be posed as the variational problem
\begin{equation}
\label{eq:ocp}
J
=
\min_{\ug(\cdot),\,L_{\mathrm{tot}}}\ \int_0^{L_{\mathrm{tot}}} 1\,ds,
\end{equation}
subject to the Sabban frame dynamics in \eqref{eq:sabban}, which are equivalent to the left invariant system in \eqref{eq:lie}, and the control bound
\begin{equation}
\label{eq:ocp_bound}
|\ug(s)|\le \Umax,
\end{equation}
and the boundary conditions
\begin{equation}
\label{eq:ocp_bc}
g(0)=R_0,\qquad g(L_{\mathrm{tot}})=R_f.
\end{equation}


\section{Numerical Solution via Least Squares}
This section describes a purely numerical method for computing shortest spherical Dubins paths using the candidate structure established in Section~II.
The objective is to robustly compute feasible candidates and select the shortest one by solving a small number of constrained nonlinear least squares problems on $\SO(3)$.

\subsection{Exponential map propagation and Rodrigues formula}
In numerical computation, the differential equation \eqref{eq:lie} must be discretized on a grid $s_k$ with step size $\Delta s$.
A direct starting point is an additive Euler update obtained by approximating $dg/ds$ with a finite difference,
\begin{equation}
\label{eq:additive_euler_so3}
g_{k+1}\approx g_k+\Delta s\,g_k\,\Omega(u_{g,k}),
\end{equation}
where $u_{g,k}$ denotes the value of the control applied on the step from $s_k$ to $s_{k+1}$ and $\Omega(u_{g,k})\in\so(3)$ is the corresponding generator.
This additive update is simple, but it does not preserve the defining constraints of $\SO(3)$.
Even if $g_k\in\SO(3)$, the matrix $g_{k+1}$ produced by \eqref{eq:additive_euler_so3} is generally not orthogonal and its determinant can deviate from one.
A numerical pipeline can attempt to repair this drift by projecting $g_{k+1}$ back to $\SO(3)$ after each step, but this introduces an additional approximation that is not tied to the underlying dynamics and can contaminate the rotation error used in the least squares objective.
These issues motivate using an update that respects the Lie group geometry of the state space and interprets the local generator $\Omega(u_{g,k})$ as an element of the tangent algebra.
Directly integrating \eqref{eq:lie} with an additive method does not preserve orthogonality and can drift off $\SO(3)$, which introduces ambiguity in interpreting a numerical state as a physical frame.
Instead, we propagate the frame using the exponential map.
Over an interval where $\ug$ is constant, \eqref{eq:g_closed_form} gives
\begin{equation}
\label{eq:num_segment_exp}
g(s)=g(0)\exp\!\left(s\,\Omega(\ug)\right),
\end{equation}
where $\exp$ denotes the matrix exponential on $\SO(3)$.
For numerical integration with a varying control, an exponential Euler step updates
\begin{equation}
\label{eq:exp_euler}
g_{k+1}=g_k\,\exp\!\left(\Delta s\,\Omega(u_{g,k})\right),
\end{equation}
which keeps $g_k$ on $\SO(3)$ up to floating point error.
To compute the matrix exponential efficiently, we use the axis angle Rodrigues formula: for a unit axis $\vct{k}$ and angle $\theta$,
\begin{equation}
\label{eq:rodrigues}
\exp\!\left(\theta\,\skewmat{\vct{k}}\right)=\mat{I}_3 + \sin\theta\,\skewmat{\vct{k}} + (1-\cos\theta)\,\skewmat{\vct{k}}^{\,2},
\end{equation}
where $\skewmat{\vct{k}}$ denotes the skew symmetric matrix corresponding to the cross product operator $\vct{v}\mapsto \vct{k}\times\vct{v}$.
This formula is not merely a convenient implementation detail.
It follows from the structure of the Lie algebra $\so(3)$ and provides a numerically reliable way to evaluate the exponential map.
For any $\vct{a}\in\R^3$, the skew symmetric matrix $\skewmat{\vct{a}}$ represents the linear operator $\vct{v}\mapsto \vct{a}\times\vct{v}$.
The exponential map from $\so(3)$ to $\SO(3)$ is defined by the matrix power series
\begin{equation}
\label{eq:exp_series}
\exp(M)=\sum_{n=0}^{\infty}\frac{M^n}{n!},
\end{equation}
which converges for all $M$.
When $M=\theta\,\skewmat{\vct{k}}$ with $\|\vct{k}\|=1$, the powers of $\skewmat{\vct{k}}$ collapse to a three dimensional span because
\begin{equation}
\label{eq:skew_powers}
\skewmat{\vct{k}}^{\,3}=-\skewmat{\vct{k}},\qquad
\skewmat{\vct{k}}^{\,4}=-\skewmat{\vct{k}}^{\,2}.
\end{equation}
Grouping the even and odd terms in the series then yields the closed form in \eqref{eq:rodrigues}, with coefficients that depend only on $\sin\theta$ and $\cos\theta$.
This reduction matters in computation for two reasons.
First, it avoids coordinate singularities and ambiguity associated with local attitude parameterizations, since the update is performed directly on $\SO(3)$.
Second, it is stable and efficient inside the least squares loop: each segment update requires only a few matrix multiplications and evaluations of $\sin\theta$ and $\cos\theta$, and the result remains an orthogonal matrix with determinant one up to floating point roundoff.
This ensures that the residual constructed through the $\SO(3)$ logarithm measures a true rotation error, rather than an artifact of numerical drift.

\subsection{Candidate set from bang bang structure}
Section~II shows that an optimal control satisfies $\ug(s)\in\{-\Umax,0,\Umax\}$.
Thus an optimal trajectory is a concatenation of tight turns $C$ with $\ug=\pm\Umax$ and geodesics $G$ with $\ug=0$.
We adopt the sign convention implied by the Sabban frame coordinates used throughout this paper: $\ug=+\Umax$ denotes a right tight turn, while $\ug=-\Umax$ denotes a left tight turn.
In regimes such as $\Rturn\in(0,1/2]$, candidate optimal path types are $CGC$, $CCC$, and degeneracies.
We therefore enumerate a finite candidate set by choosing the turn directions for each $C$ segment and assigning $\ug=0$ for each $G$ segment.

\subsection{Constrained least squares on $\SO(3)$}
For a fixed candidate path type, the unknowns are only the segment lengths.
In principle, one could attempt to solve the boundary equation $g(L_{\mathrm{tot}})=R_f$ directly as a system of nonlinear equations.
In practice, however, numerical solvers benefit from a smooth objective that quantifies how far a tentative set of segment lengths is from satisfying the boundary condition.
This is particularly important here because the mapping from segment lengths to the terminal frame is nonlinear, periodic, and can admit multiple feasible solutions with different lengths.
We therefore pose boundary matching as a constrained least squares problem on $\SO(3)$.
The key idea is to represent the rotation mismatch by a tangent space residual using the logarithm map.
This produces a three dimensional residual vector that varies smoothly with the segment lengths near a feasible solution, so standard trust region least squares methods can be applied reliably.
Once a candidate produces a small residual, the corresponding segment lengths define a feasible path, and the shortest feasible path is obtained by comparing total length across candidates.

To keep the expressions compact when $g(0)=R_0$, we work with the normalized frame $\tilde{g}(s)=R_0^{\T}g(s)$.
Then $\tilde{g}(0)=\mat{I}_3$ and the boundary condition becomes $\tilde{g}(L_{\mathrm{tot}})=R$, where $R$ denotes the normalized target rotation $R=R_0^{\T}R_f$.
For a fixed candidate with piecewise constant controls $(u_{g,1},u_{g,2},u_{g,3})$ and nonnegative segment lengths $(s_1,s_2,s_3)$, the normalized terminal frame is
\begin{equation}
\label{eq:prod}
\tilde{g}(L_{\mathrm{tot}})=\exp(\Omega(u_{g,1})s_1)\,\exp(\Omega(u_{g,2})s_2)\,\exp(\Omega(u_{g,3})s_3),
\end{equation}
where $\exp$ denotes the matrix exponential, and feasibility is expressed by matching $\tilde{g}(L_{\mathrm{tot}})=R$.
We measure the mismatch using the matrix logarithm from $\SO(3)$ to $\so(3)$.
In terms of the original frame, the corresponding terminal rotation is $g(L_{\mathrm{tot}})=R_0\,\tilde{g}(L_{\mathrm{tot}})$.
\subsubsection{Residual definition for the boundary matching problem}
The feasibility condition $\tilde{g}(L_{\mathrm{tot}})=R$ is equivalent to the root finding problem $\vct{r}(\vct{s})=\vct{0}$, where we collect the unknown segment lengths in $\vct{s}=[s_1,s_2,s_3]^{\T}$.
The residual is defined through the relative rotation
\begin{equation}
\label{eq:so3_residual_def}
R_{\mathrm{err}}(\vct{s})=R^{\T}\tilde{g}(L_{\mathrm{tot}};\vct{s}),\qquad
\skewmat{\vct{r}(\vct{s})}=\log\!\left(R_{\mathrm{err}}(\vct{s})\right),
\end{equation}
where $\log$ denotes the matrix logarithm and $\skewmat{\cdot}$ maps a vector to its corresponding skew symmetric matrix, which yields a three dimensional vector $\vct{r}(\vct{s})\in\R^3$ measuring the axis angle error.
For numerical evaluation, the logarithm can be computed in closed form.
The vee map extracts the vector associated with a skew symmetric matrix.
Concretely, for any $\vct{v}\in\R^3$, the vee map applied to $\skewmat{\vct{v}}$ returns $\vct{v}$, and for $S\in\so(3)$ we have
\begin{equation}
\label{eq:vee_components}
\text{vee}(S)=\begin{bmatrix} S_{32} \\ S_{13} \\ S_{21} \end{bmatrix},
\end{equation}
where $\text{vee}(\cdot)$ denotes the vee map operation.
Let $\theta$ denote the rotation angle obtained from the trace,
\begin{equation}
\label{eq:omega_from_Rerr}
\theta=\cos^{-1}\!\left(\frac{\trace(R_{\mathrm{err}}(\vct{s}))-1}{2}\right),\qquad
\vct{r}(\vct{s})=\frac{\theta}{2\sin\theta}\,\text{vee}\!\left(R_{\mathrm{err}}(\vct{s})-R_{\mathrm{err}}(\vct{s})^{\T}\right),
\end{equation}
with the limiting approximation $\vct{r}(\vct{s})\approx \text{vee}\!\left((R_{\mathrm{err}}(\vct{s})-R_{\mathrm{err}}(\vct{s})^{\T})/2\right)$ when $\theta$ is close to zero.

\subsubsection{Gauss Newton iterations for the least squares solve}
We compute $\vct{s}$ by Gauss Newton iterations.
Starting from an initial guess $\vct{s}_0$ within the admissible bounds, we generate a sequence $\{\vct{s}_k\}$ by repeatedly evaluating the residual $\vct{r}(\vct{s}_k)$, building an approximate Jacobian $J(\vct{s}_k)$, solving for an update direction $\bm{\delta}_k$, and then updating $\vct{s}_{k+1}$.
In practice, we use a small deterministic set of initial guesses and select the seed with the smallest initial residual norm before iterating.
The update $\bm{\delta}$ is computed by solving the normal equations
\begin{equation}
\label{eq:gn_normal}
J(\vct{s})^{\T}J(\vct{s})\,\bm{\delta}=-J(\vct{s})^{\T}\vct{r}(\vct{s}),
\end{equation}
followed by the parameter update $\vct{s}\leftarrow \vct{s}+\bm{\delta}$.
In our setting, $J(\vct{s})\in\R^{3\times 3}$ is approximated by finite differences.
Let $\vct{e}_j$ denote the $j$th standard basis vector in $\R^3$ so that $(\vct{e}_j)_\ell=1$ if $\ell=j$ and $(\vct{e}_j)_\ell=0$ otherwise.
Let $h_j>0$ denote a step size for perturbing the $j$th segment length $s_j$ when forming difference quotients.
Write $\vct{r}(\vct{s})=[r_1(\vct{s}),r_2(\vct{s}),r_3(\vct{s})]^{\T}$.
Then $J(\vct{s})_{ij}=\partial r_i/\partial s_j$ denotes the derivative of the $i$th residual component with respect to the $j$th segment length, where $i\in\{1,2,3\}$ and $j\in\{1,2,3\}$ correspond to the three residual components and the three segment lengths $(s_1,s_2,s_3)$.
An entrywise finite difference approximation is
\begin{equation}
\label{eq:fd_jacobian_entries}
J(\vct{s})_{ij}\approx \frac{r_i(\vct{s}+h_j\vct{e}_j)-r_i(\vct{s}-h_j\vct{e}_j)}{2h_j},
\end{equation}
and equivalently the $j$th column can be written as
\begin{equation}
\label{eq:fd_jacobian}
J(\vct{s})\vct{e}_j\approx \frac{\vct{r}(\vct{s}+h_j\vct{e}_j)-\vct{r}(\vct{s}-h_j\vct{e}_j)}{2h_j}.
\end{equation}
In practice we select $h_j$ relative to the scale of $s_j$, for example $h_j=\eta\max(1,|s_j|)$ with a small $\eta$.
After the update, we enforce the bounds by projecting each component of $\vct{s}$ into its admissible interval using componentwise clamping: for the $i$th component, if the updated value $(s_k+\bm{\delta}_k)_i$ falls outside the bounds $[0, b_i]$, it is clamped to the nearest bound, where $b_i=2\pi$ for $G$ segments and $b_i=2\pi\Rturn$ for $C$ segments.
The iteration terminates when $\|\vct{r}(\vct{s})\|_2$ falls below a tolerance, or when the update becomes small, $\|\bm{\delta}\|_2\le \varepsilon_{\delta}$.

Because each constant control segment is periodic on the sphere, we also bound each $s_i$ by one period of its segment type to reduce redundant minima.
For a $G$ segment the period is $2\pi$, while for a $C$ segment with radius $\Rturn$ the period is $2\pi\Rturn$.
For clarity, the numerical procedure can be summarized as follows:

\begin{algorithm}[htbp]
\caption{Gauss-Newton Least Squares on $\SO(3)$}
\label{alg:gn_so3}
\begin{algorithmic}[1]
\State \textbf{Input:} Candidate path type $(u_{g,1},u_{g,2},u_{g,3})\in\{-\Umax,0,\Umax\}^3$, normalized target rotation $R$, tolerance $\varepsilon$
\State \textbf{Output:} Segment lengths $\vct{s}=[s_1,s_2,s_3]^{\T}$ or failure indicator
\State
\State Choose bounds for segment lengths: for $G$ segments, $0\le s_i\le 2\pi$; for $C$ segments, $0\le s_i\le 2\pi\Rturn$
\State Choose initial guess $\vct{s}_0=[s_1,s_2,s_3]^{\T}$ within bounds (e.g., from deterministic seed set, selecting best by residual norm)
\State
\For{$k=0,1,2,\ldots$}
    \State Compute residual $\vct{r}(\vct{s}_k)$ using \eqref{eq:so3_residual_def} and \eqref{eq:omega_from_Rerr}
    \If{$\|\vct{r}(\vct{s}_k)\|_2 < \varepsilon$}
        \State \Return $\vct{s}_k$ \Comment{Converged}
    \EndIf
    \State Approximate Jacobian $J(\vct{s}_k)$ by finite differences \eqref{eq:fd_jacobian} with steps $h_j$
    \State Solve normal equations \eqref{eq:gn_normal} for $\bm{\delta}_k$
    \If{$\|\bm{\delta}_k\|_2 < \varepsilon_{\delta}$}
        \State \Return $\vct{s}_k$ \Comment{Update too small}
    \EndIf
    \State Update $\vct{s}_{k+1}=\Pi(\vct{s}_k+\bm{\delta}_k)$, where $\Pi$ clamps each component $(s_{k+1})_i$ to $[0, b_i]$ with $b_i=2\pi$ for $G$ segments and $b_i=2\pi\Rturn$ for $C$ segments
\EndFor
\State
\State After solving each candidate, select feasible candidate with minimal total length $L_{\mathrm{tot}}=s_1+s_2+s_3$
\end{algorithmic}
\end{algorithm}

The least squares procedure above is local in nature.
For a fixed candidate path type, the map from segment lengths to the terminal frame is nonlinear and periodic, and a single boundary configuration can admit multiple feasible solutions with different total lengths.
As a result, the computed solution can depend on the initial guess and, without an exhaustive multi start search, feasible candidates can be missed.
These limitations motivate an analytic method that solves the boundary constraint more directly by exploiting the algebraic structure of the exponential product, as developed in Section~IV.

\section{Analytic Solution Method}
\label{sec:semi_analytic}
This section develops an analytic solution viewpoint for spherical Dubins candidates based on Lie group factorization on $\SO(3)$.
We show how constant control segments can be represented as elementary rotations, and how boundary matching can be reformulated as a low dimensional algebraic reconstruction problem once a candidate path type is fixed.
The presentation focuses on the CGC family and outlines the corresponding ingredients needed to treat CCC candidates.

\subsection{Generator matrices and axis angle interpretation}
Throughout this section we work with the normalized boundary value problem.
Given initial and terminal frames $g(0)=R_0$ and $g(L_{\mathrm{tot}})=R_f$, define the normalized target $R=R_0^{\T}R_f$ and the normalized trajectory $\tilde{g}(s)=R_0^{\T}g(s)$, so that $\tilde{g}(0)=\mat{I}_3$ and $\tilde{g}(L_{\mathrm{tot}})=R$.
For notational simplicity we drop the tilde and write $g(0)=\mat{I}_3$ and $g(L_{\mathrm{tot}})=R$ in what follows.
Once a candidate path is fixed, each segment corresponds to a constant generator and therefore to an elementary rotation.
The boundary condition becomes an exponential product equation on $\SO(3)$, directly analogous to the forward kinematics of a serial mechanism in robotics.
The unknown segment lengths play the role of joint variables, and solving the boundary constraint is an inverse kinematics problem.
In favorable cases, this inverse problem can be reduced to a small number of scalar constraints with closed form reconstruction of the remaining parameters, which yields multiple solutions without a multi start search.
To keep the exposition concrete, we adopt the explicit matrix form \eqref{eq:Omega_matrix} for the generator in the left invariant system \eqref{eq:lie}.
For constant $\ug\equiv u$, the solution over an interval of arc length $s$ is
\begin{equation}
\label{eq:segment_exp}
g(s)=g(0)\exp\!\left(s\,\Omega(u)\right).
\end{equation}
Moreover, $\exp(s\,\Omega(u))$ is a rotation with an axis angle representation.
In the coordinate convention implied by \eqref{eq:Omega_matrix}, the associated body fixed angular velocity vector is proportional to
\begin{equation}
\label{eq:axis_for_u}
\vct{w}(u)=\begin{bmatrix} u \\ 0 \\ 1 \end{bmatrix},\qquad
\vct{k}(u)=\frac{\vct{w}(u)}{\|\vct{w}(u)\|},
\end{equation}
so that $\exp(s\,\Omega(u))$ can be interpreted as a rotation by angle $\theta=s\|\vct{w}(u)\|$ about axis $\vct{k}(u)$.
We denote this rotation compactly by $R(\vct{k},\theta)$ and evaluate it using the Rodrigues formula in \eqref{eq:rodrigues}.
For convenience, the Rodrigues formula can be expanded entrywise.
Let $\vct{k}=[k_1,k_2,k_3]^{\T}$ with $\|\vct{k}\|=1$, $c=\cos\theta$, and $s=\sin\theta$.
Then
\begin{equation}
\label{eq:rodrigues_entries}
R(\vct{k},\theta)=
\begin{bmatrix}
c+k_1^2(1-c) & k_1k_2(1-c)-k_3 s & k_1k_3(1-c)+k_2 s\\
k_2k_1(1-c)+k_3 s & c+k_2^2(1-c) & k_2k_3(1-c)-k_1 s\\
k_3k_1(1-c)-k_2 s & k_3k_2(1-c)+k_1 s & c+k_3^2(1-c)
\end{bmatrix}.
\end{equation}
In particular, for tight turns $u=\pm \Umax$, we have $\|\vct{w}(\pm\Umax)\|=\sqrt{1+\Umax^2}=1/\Rturn$ and thus
\begin{equation}
\label{eq:theta_s_relation}
\theta=\frac{s}{\Rturn}\quad\text{for}\quad u=\pm\Umax,\qquad
\theta=s\quad\text{for}\quad u=0.
\end{equation}

\subsection{Exponential product representation}
For a piecewise constant control sequence $(u_{g,1},u_{g,2},u_{g,3})$ with segment lengths $(s_1,s_2,s_3)$, the terminal frame admits the exponential product representation \eqref{eq:prod}, and the boundary condition becomes $g(L_{\mathrm{tot}})=R$.
Thus, solving \eqref{eq:ocp} reduces to enumerating candidate control path types and solving a finite dimensional equation on $\SO(3)$ for the segment lengths.

\subsection{Analytic solver core result}
Our analytic solver is best understood through a robot kinematics lens.
Equation \eqref{eq:prod} is an \emph{exponential product} on $\SO(3)$, directly analogous to the forward kinematics of a serial manipulator written as a product of elementary joint motions.
Here each segment type ($C$ or $G$) plays the role of a joint with a \emph{known rotation axis} and an \emph{unknown joint displacement} (the segment angle/length).
Thus the boundary condition $g(L_{\mathrm{tot}})=R$ becomes an inverse kinematics problem that solves for the segment parameters that realize a desired end frame $R$.

The key insight is that both CGC and CCC path types follow the same elimination strategy: by exploiting the axis-fixing property of rotations, we reduce the three-dimensional boundary constraint to a single scalar equation in $\theta_1$, then recover the remaining parameters $(\theta_2,\theta_3)$ via closed-form back substitution.
The only difference between CGC and CCC lies in the specific coefficients $\alpha$, $\beta$, and $\gamma$ that appear in the trigonometric constraint equation.

\subsubsection{Feasibility lemmas}
Before presenting the factorization strategy, we state two lemmas that are central to the elimination procedure.

\begin{lemma}
\label{lem:axis_feasible}
Let $\vct{k}\in\Sph^2$ and let $\vct{v},\vct{w}\in\Sph^2$.
There exists a rotation angle $\theta$ such that $R(\vct{k},\theta)\vct{v}=\vct{w}$ if and only if
\begin{equation}
\label{eq:dot_invariant}
\vct{k}^{\T}\vct{v}=\vct{k}^{\T}\vct{w}.
\end{equation}
\end{lemma}

\begin{lemma}
\label{lem:axis_fixed}
Let $Q\in\SO(3)$ and let $\vct{k}\in\Sph^2$.
Then $Q$ is a rotation about axis $\vct{k}$, i.e., $Q=R(\vct{k},\theta)$ for some $\theta$, if and only if $Q\vct{k}=\vct{k}$.
Equivalently,
\begin{equation}
\label{eq:axis_fixed_scalar}
\vct{k}^{\T}Q\vct{k}=1.
\end{equation}
\end{lemma}

\begin{figure}[htbp]
\centering
\begin{subfigure}[b]{0.48\textwidth}
\centering
\includegraphics[width=\textwidth]{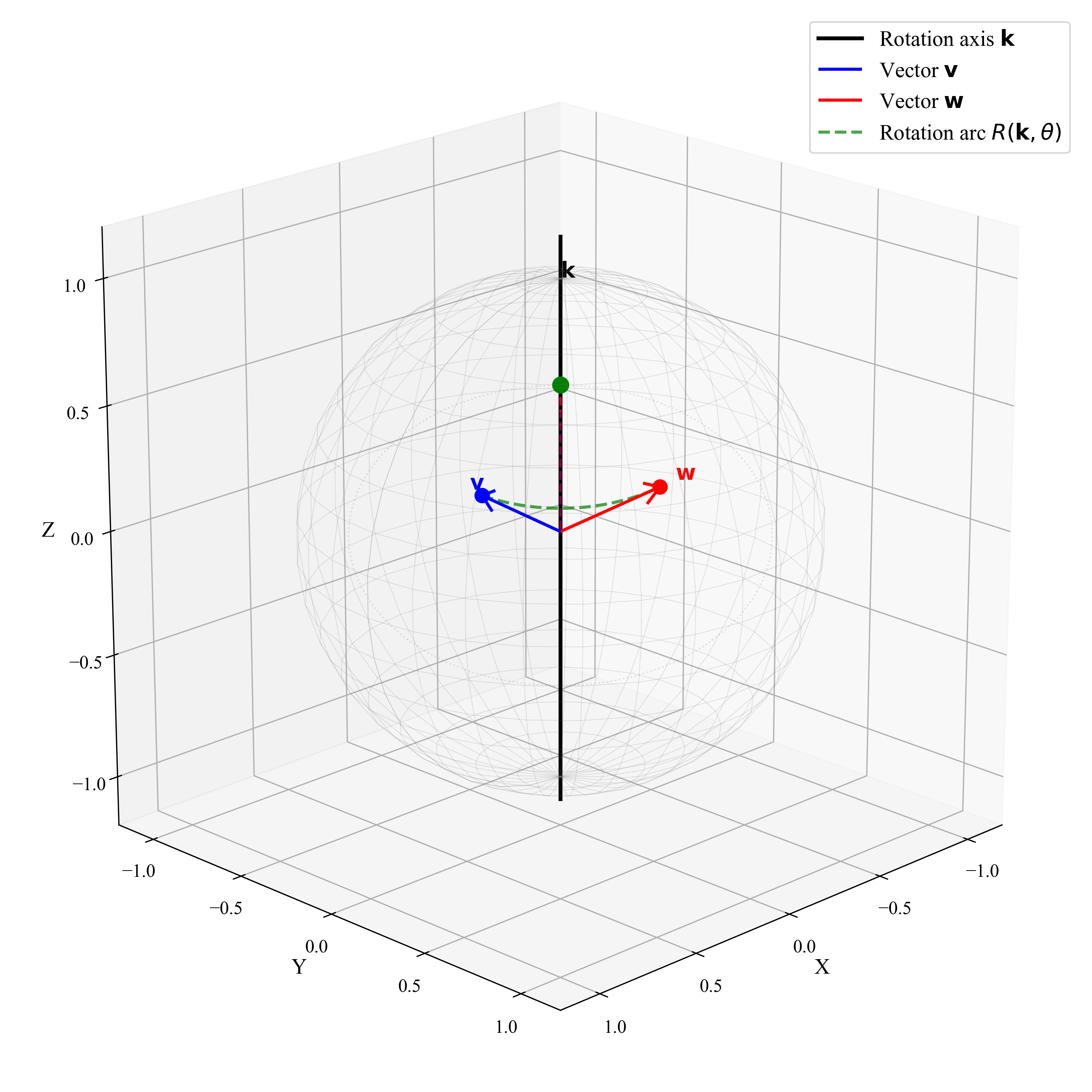}
\caption{Lemma~\ref{lem:axis_feasible}: rotation feasibility condition}
\label{fig:lemma1}
\end{subfigure}
\hfill
\begin{subfigure}[b]{0.48\textwidth}
\centering
\includegraphics[width=\textwidth]{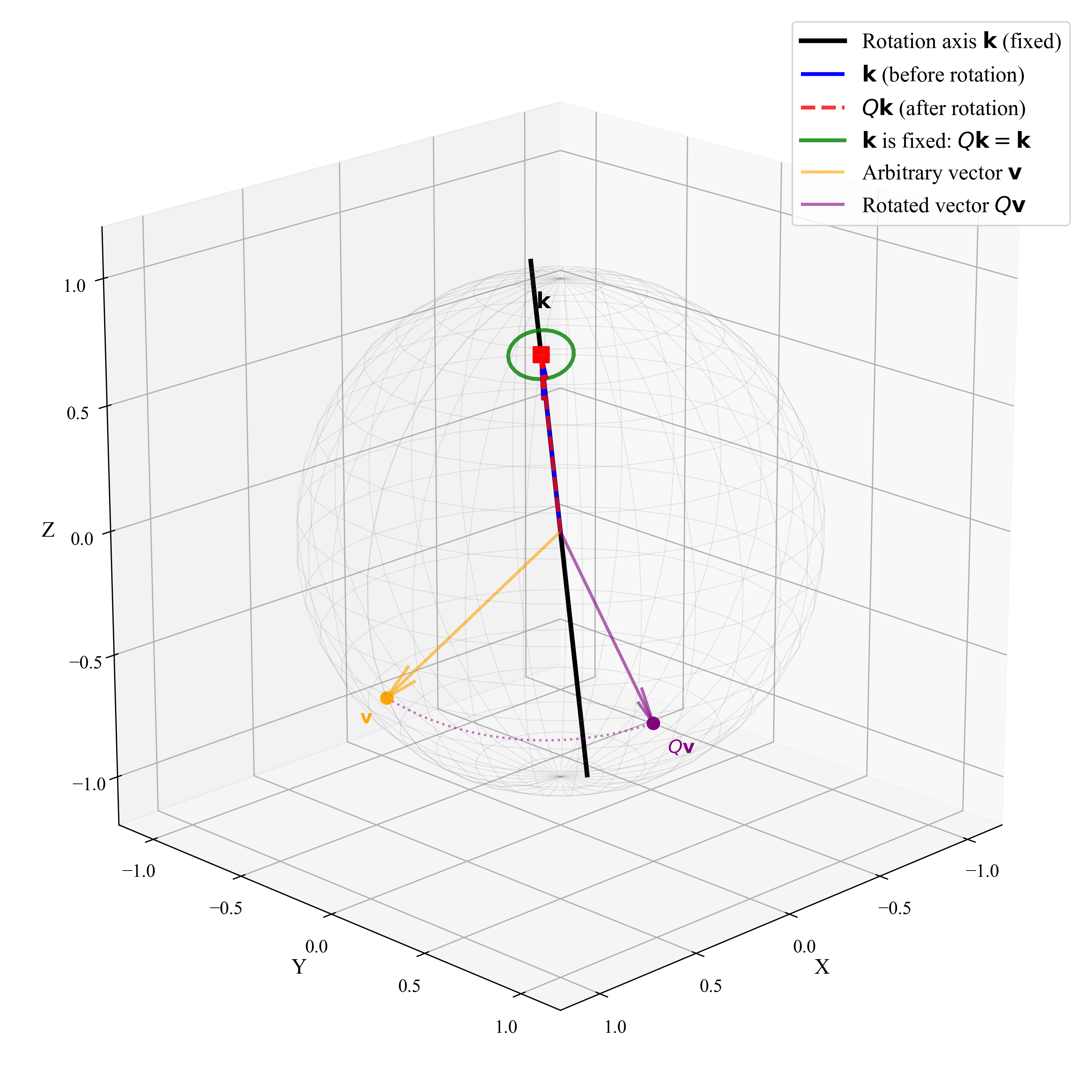}
\caption{Lemma~\ref{lem:axis_fixed}: axis-fixing property}
\label{fig:lemma2}
\end{subfigure}
\caption{Geometric illustration of the feasibility lemmas.
Lemma~\ref{lem:axis_feasible} shows that a rotation $R(\vct{k},\theta)$ mapping $\vct{v}$ to $\vct{w}$ exists if and only if the projections of $\vct{v}$ and $\vct{w}$ onto the rotation axis $\vct{k}$ are equal.
Lemma~\ref{lem:axis_fixed} shows that a rotation matrix $Q$ is a rotation about axis $\vct{k}$ if and only if $Q$ fixes $\vct{k}$, i.e., $Q\vct{k}=\vct{k}$.}
\label{fig:lemmas}
\end{figure}

\subsubsection{Elimination of $\theta_2$}
Consider a fixed candidate path with controls $(u_1,u_2,u_3)$ where each $u_i\in\{0,\pm\Umax\}$.
For CGC path types, $u_2=0$ (geodesic middle segment), while for CCC path types, $u_2=\pm\Umax$ (tight turn middle segment).
Using \eqref{eq:axis_for_u} to \eqref{eq:theta_s_relation}, define axes $\vct{k}_i=\vct{k}(u_i)$ for $i=1,2,3$ and angles $\theta_i=s_i/\Rturn$ for tight turns ($u_i=\pm\Umax$) or $\theta_i=s_i$ for geodesics ($u_i=0$).

Let
\begin{equation}
\label{eq:R123_def}
\mat{R}_1(\theta_1)=R(\vct{k}_1,\theta_1),\qquad
\mat{R}_2(\theta_2)=R(\vct{k}_2,\theta_2),\qquad
\mat{R}_3(\theta_3)=R(\vct{k}_3,\theta_3),
\end{equation}
where for CGC path types, $\vct{k}_2=\ethree=[0,0,1]^{\T}$ since $u_2=0$ generates a rotation about the $z$ axis.
The boundary condition in normalized coordinates, $\tilde{g}(L_{\mathrm{tot}})=R$ with $R=R_0^{\T}R_f$, becomes
\begin{equation}
\label{eq:target}
R=\mat{R}_1(\theta_1)\mat{R}_2(\theta_2)\mat{R}_3(\theta_3).
\end{equation}
Rearranging,
\begin{equation}
\label{eq:R2_residual}
\mat{R}_2(\theta_2)=\mat{R}_1(\theta_1)^{\T}\,R\,\mat{R}_3(\theta_3)^{\T}.
\end{equation}
Since $\mat{R}_2(\theta_2)$ is a rotation about axis $\vct{k}_2$, by Lemma~\ref{lem:axis_fixed} it must fix $\vct{k}_2$, i.e., $\mat{R}_2(\theta_2)\vct{k}_2=\vct{k}_2$.
Applying this to \eqref{eq:R2_residual} yields
\begin{equation}
\label{eq:k2_fixed}
\vct{k}_2=\mat{R}_1(\theta_1)^{\T}\,R\,\mat{R}_3(\theta_3)^{\T}\vct{k}_2.
\end{equation}
Multiplying both sides by $\mat{R}_1(\theta_1)$ and rearranging gives
\begin{equation}
\label{eq:R3Tk2}
\mat{R}_3(\theta_3)^{\T}\vct{k}_2=R^{\T}\mat{R}_1(\theta_1)\vct{k}_2.
\end{equation}

\subsubsection{Elimination of $\theta_3$}
To derive a scalar feasibility condition from \eqref{eq:R3Tk2}, left multiply both sides by $\vct{k}_3^{\T}$.
Since $\mat{R}_3(\theta_3)$ is a rotation about axis $\vct{k}_3$, the component along $\vct{k}_3$ is invariant, so $\vct{k}_3^{\T}\mat{R}_3(\theta_3)^{\T}\vct{k}_2=\vct{k}_3^{\T}\vct{k}_2$.
Applying this invariance to \eqref{eq:R3Tk2} yields the dot product constraint
\begin{equation}
\label{eq:scalar_condition}
\vct{k}_3^{\T}\left(R^{\T}\mat{R}_1(\theta_1)\vct{k}_2\right)=\vct{k}_3^{\T}\vct{k}_2.
\end{equation}
By Lemma~\ref{lem:axis_feasible}, \eqref{eq:R3Tk2} admits a solution $\theta_3$ if and only if this dot product invariance holds, which yields a scalar equation in the single unknown $\theta_1$.

Expanding \eqref{eq:scalar_condition} using the Rodrigues formula \eqref{eq:rodrigues_entries} and collecting terms, we obtain a trigonometric equation linear in $\cos\theta_1$ and $\sin\theta_1$:
\begin{equation}
\label{eq:trig_scalar}
\alpha+\beta\cos\theta_1+\gamma\sin\theta_1=0,
\end{equation}
where the coefficients $\alpha$, $\beta$, and $\gamma$ depend on the entries of $R$ and the specific path type (CGC or CCC).
The key difference between CGC and CCC lies solely in these coefficient expressions, while the elimination and back-substitution procedures are identical.
\subsubsection{Solution for $\theta_1$, $\theta_2$, $\theta_3$}
Given coefficients $\alpha$, $\beta$, and $\gamma$ from \eqref{eq:trig_scalar}, we solve for $\theta_1$.
For both CGC and CCC path types, the half-angle substitution $t=\tan(\theta_1/2)$ converts \eqref{eq:trig_scalar} into a quadratic polynomial:
\begin{equation}
\label{eq:half_angle_quadratic}
(\alpha-\beta)t^2+2\gamma t+(\alpha+\beta)=0,
\end{equation}
which can be solved in closed form, typically yielding up to two roots in $[0,2\pi)$.

Once $\theta_1$ is found, we recover $\theta_3$ and $\theta_2$ via closed-form back substitution.
Define the target vector
\begin{equation}
\label{eq:w_target}
\vct{w}(\theta_1)=R^{\T}\mat{R}_1(\theta_1)\vct{k}_2.
\end{equation}
The mapping equation \eqref{eq:R3Tk2} is $\mat{R}_3(\theta_3)^{\T}\vct{k}_2=\vct{w}(\theta_1)$, where $\mat{R}_3(\theta_3)$ is a rotation about axis $\vct{k}_3$.
When \eqref{eq:scalar_condition} holds, Lemma~\ref{lem:axis_feasible} guarantees existence of such a $\theta_3$.
To compute it explicitly, project $\vct{k}_2$ and $\vct{w}$ onto the plane orthogonal to $\vct{k}_3$:
\begin{equation}
\label{eq:proj_plane}
\vct{v}_0=\vct{k}_2-\vct{k}_3(\vct{k}_3^{\T}\vct{k}_2),\qquad
\vct{v}=\vct{w}(\theta_1)-\vct{k}_3(\vct{k}_3^{\T}\vct{w}(\theta_1)).
\end{equation}
If $\|\vct{v}_0\|$ and $\|\vct{v}\|$ are nonzero, normalize and compute the signed in-plane rotation angle
\begin{equation}
\label{eq:theta3_atan2}
\phi=\atanTwo\!\Big(\vct{k}_3^{\T}\big(\hat{\vct{v}}_0\times \hat{\vct{v}}\big),\ \hat{\vct{v}}_0^{\T}\hat{\vct{v}}\Big),
\end{equation}
where $\hat{\vct{v}}_0=\vct{v}_0/\|\vct{v}_0\|$ and $\hat{\vct{v}}=\vct{v}/\|\vct{v}\|$.
Since \eqref{eq:R3Tk2} involves $\mat{R}_3(\theta_3)^{\T}=R(\vct{k}_3,-\theta_3)$, we take $\theta_3=-\phi$ and wrap to $[0,2\pi)$.

With $(\theta_1,\theta_3)$ available, form the residual rotation
\begin{equation}
\label{eq:Rres_for_theta2}
\mat{R}_{\mathrm{res}}=\mat{R}_1(\theta_1)^{\T}R\mat{R}_3(\theta_3)^{\T}.
\end{equation}
For an exact solution, $\mat{R}_{\mathrm{res}}$ equals $\mat{R}_2(\theta_2)=R(\vct{k}_2,\theta_2)$ and is therefore a rotation about axis $\vct{k}_2$.
For CGC path types where $\vct{k}_2=\ethree$, $\theta_2$ is obtained directly from the $(1,1)$ and $(2,1)$ entries as $\theta_2=\atanTwo((\mat{R}_{\mathrm{res}})_{21},(\mat{R}_{\mathrm{res}})_{11})$.
For CCC path types, $\theta_2$ is extracted similarly by projecting onto the plane orthogonal to $\vct{k}_2$ and using the same atan2 procedure.

Finally, recover segment lengths: for tight turns ($u_i=\pm\Umax$), $s_i=\Rturn\theta_i$; for geodesics ($u_i=0$), $s_i=\theta_i$.

\subsubsection{Coefficient expressions for representative path types}
The coefficients $\alpha$, $\beta$, and $\gamma$ in \eqref{eq:trig_scalar} depend on the specific path type.
For the CGC path $RGL$ with $(u_1,u_2,u_3)=(+\Umax,0,-\Umax)$, the coefficients are
\begin{equation}
\label{eq:rgl_coeffs}
\begin{aligned}
\alpha_{RGL} &= -1-\Umax\big(\Umax+\Umax R_{11}-R_{13}+R_{31}\big)+R_{33},\\
\beta_{RGL}  &= \Umax\big(-R_{13}+\Umax(R_{11}-\Umax R_{31}+R_{33})\big),\\
\gamma_{RGL} &= \Umax\sqrt{1+\Umax^2}\,(\Umax R_{21}-R_{23}),
\end{aligned}
\end{equation}
where $R_{ij}$ denote the entries of the normalized target rotation $R$.

For the CCC path $RLR$ with $(u_1,u_2,u_3)=(+\Umax,-\Umax,+\Umax)$, the coefficients are
\begin{equation}
\label{eq:rlr_coeffs}
\begin{aligned}
\alpha_{RLR} &= -\big((\Umax^2-1)\big(\Umax\big((R_{11}-1)\Umax+R_{13}+R_{31}\big)+R_{33}-1\big)\big),\\
\beta_{RLR}  &= 2\Umax\big(\Umax(-R_{11}+R_{31}\Umax+R_{33})-R_{13}\big),\\
\gamma_{RLR} &= -2\Umax\sqrt{1+\Umax^2}(R_{21}\Umax+R_{23}).
\end{aligned}
\end{equation}

Coefficients for the remaining CGC and CCC path types are provided in the Appendix.

\subsubsection{Degenerate cases and completeness}
The analytic solution method developed above for CGC and CCC path types can be extended to handle their degenerate cases, which correspond to paths with fewer than three segments.
Specifically, the degenerate cases from Lemma~\ref{lem:optimal_path_types} include $CG$, $GC$, $CC$, $C$, and $G$ path types.
These can be treated as special cases of the three-segment formulation where one or more segment lengths vanish.
For instance, a $CG$ path can be viewed as a CGC path with $\theta_3=0$, while a $CC$ path corresponds to a CCC path with the middle segment length set to zero.
The two-segment cases ($CG$, $GC$, $CC$) reduce to simpler boundary value problems that can be solved using similar elimination techniques, while single-segment paths ($C$ and $G$) are trivial and correspond to single rotations.
By Lemma~\ref{lem:optimal_path_types}, when $\Rturn \in (0, 1/2]$, the optimal path must be one of the types $CGC$, $CCC$, $CG$, $GC$, $CC$, $C$, or $G$.
Since the unified analytic method provides closed-form solutions for CGC and CCC paths, and the degenerate cases can be handled through the same framework or simpler variants, the proposed approach effectively solves the optimal trajectory computation problem for the entire parameter regime $\Rturn \in (0, 1/2]$.

\section{Numerical Experiments}
We validate the proposed analytic solver through comprehensive comparisons with a baseline numerical least squares approach on $\SO(3)$. All experiments run on a laptop with an Intel 12th Gen Core i9 12900H, base frequency 2.5 GHz, 14 physical cores and 20 logical processors.In terms of software, we use Python 3.10.9 for all computations.

\subsection{Solution Completeness}
To demonstrate the completeness advantage of the analytic method, we consider a specific test case: connecting an initial configuration with position $\vct{X}_0=[1,0,0]^{\T}$ and tangent $\vct{T}_0=[0,1,0]^{\T}$ to a target configuration with position $\vct{X}_f=[0,1,0]^{\T}$ and tangent $\vct{T}_f=[0,0,1]^{\T}$.
In the normalized Sabban frame formulation, this corresponds to $R_0=\mat{I}_3$ and 
\begin{equation}
\label{eq:Rf_test_case}
R_f=\begin{bmatrix} 0 & 0 & 1 \\ 1 & 0 & 0 \\ 0 & 1 & 0 \end{bmatrix}.
\end{equation}
This represents a $90^\circ$ rotation about the axis $[1,1,1]^{\T}/\sqrt{3}$.

\begin{figure}[htbp]
\centering
\begin{subfigure}[b]{0.48\textwidth}
\centering
\includegraphics[width=\textwidth]{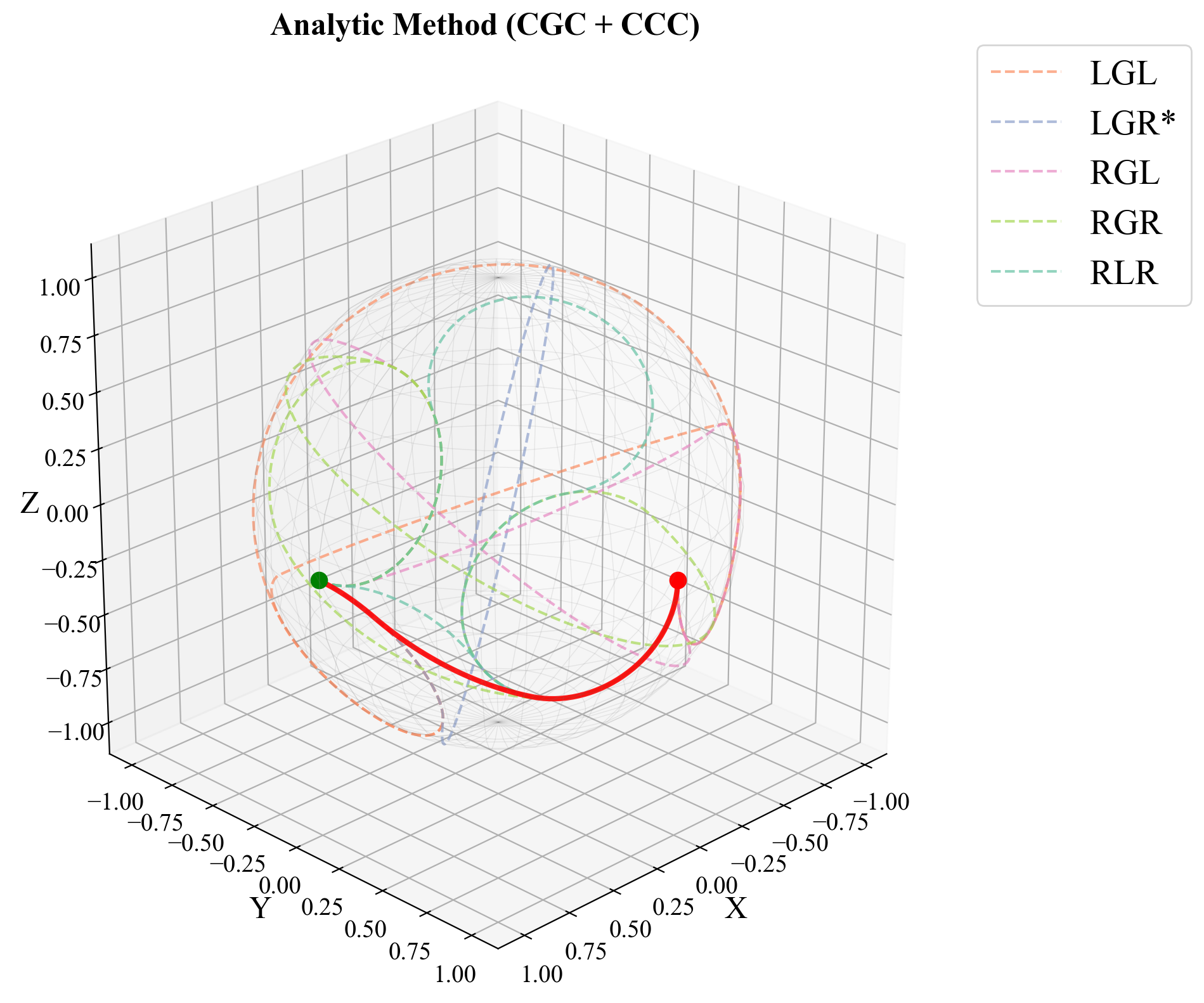}
\caption{Analytic method}
\label{fig:analytic_comparison}
\end{subfigure}
\hfill
\begin{subfigure}[b]{0.48\textwidth}
\centering
\includegraphics[width=\textwidth]{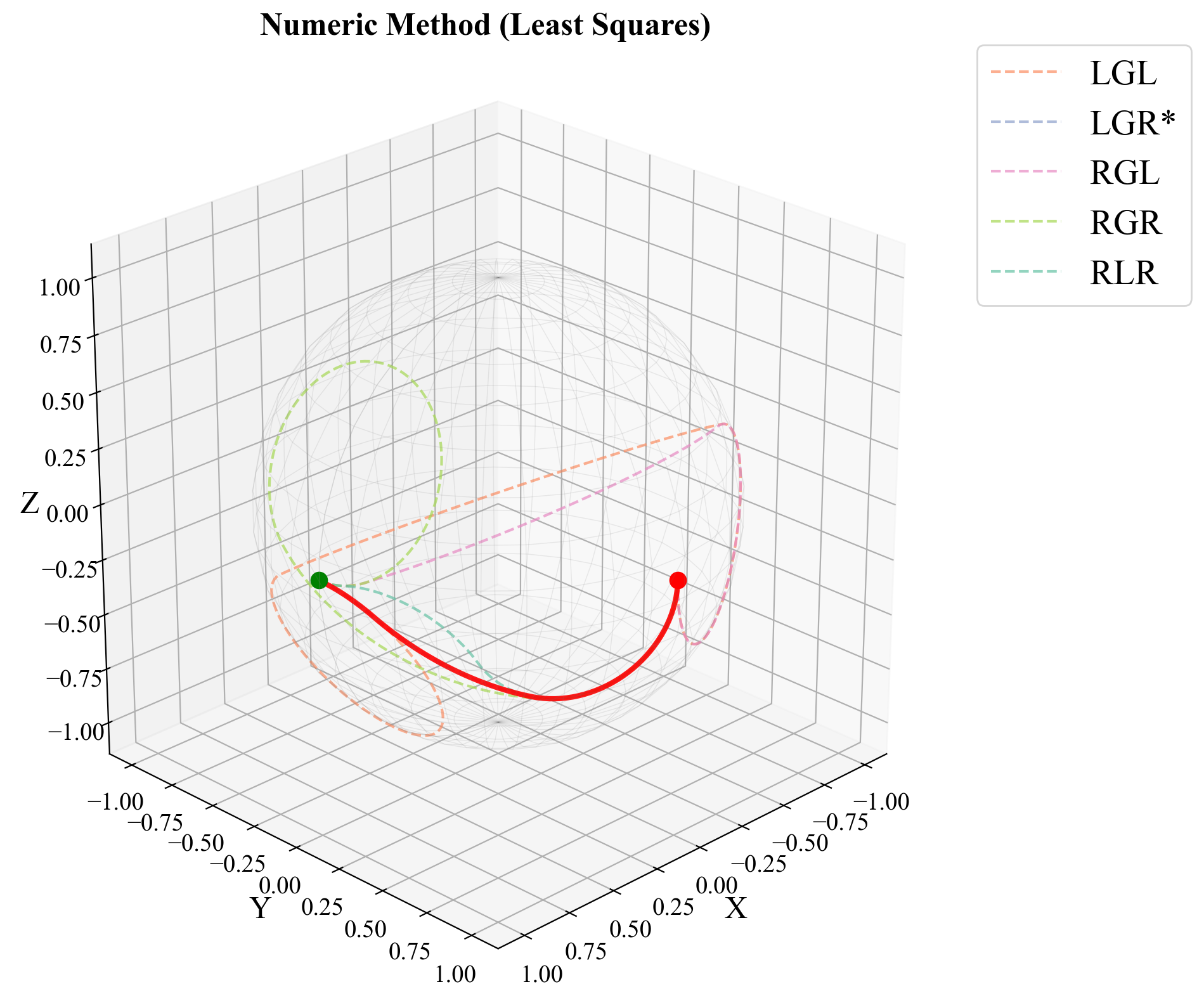}
\caption{Numeric method}
\label{fig:numeric_comparison}
\end{subfigure}
\caption{Comparison of solution completeness for the test case \eqref{eq:Rf_test_case} with $U_{\max}=2.0$.
All candidate trajectories are shown as dashed lines, with the optimal LGR path highlighted as a solid red line.
The asterisk ($*$) denotes the shortest path among all feasible solutions.
The analytic method finds all feasible solution branches, while the numeric method may miss solutions depending on initialization.}
\label{fig:solution_completeness}
\end{figure}

Figure~\ref{fig:solution_completeness} visualizes the candidate trajectories found by both methods.
The analytic solver successfully identifies all feasible solution branches across both CGC and CCC path families, as shown in Fig.~\ref{fig:analytic_comparison}.
In contrast, the numeric least squares approach, shown in Fig.~\ref{fig:numeric_comparison}, finds fewer solutions because it relies on local optimization from a limited set of initial guesses and can converge to the same solution from different starting points.

\begin{table}[htbp]
\centering
\caption{Solution comparison for test case \eqref{eq:Rf_test_case} with $U_{\max}=2.0$ and tolerance $\epsilon=10^{-13}$.
The analytic method finds all feasible branches (indicated by branch fractions $m/n$), while the numeric method finds one solution per candidate path.
Segment lengths $s_1$, $s_2$, and $s_3$ are shown for each solution.}
\label{tab:solution_comparison}
\begin{tabular}{lccccccc}
\toprule
\textbf{Candidate} & \textbf{Method} & \textbf{Length} & \textbf{Error} & \textbf{$s_1$} & \textbf{$s_2$} & \textbf{$s_3$} & \textbf{Branches} \\
\midrule
CCC(R-L-R) & Analytic & 1.965618 & $6.423\times 10^{-16}$ & 0.131140 & 0.674512 & 1.159966 & 1/2 \\
CCC(R-L-R) & Analytic & 5.318358 & $6.947\times 10^{-16}$ & 1.077060 & 2.135413 & 2.105885 & 2/2 \\
CGC(L-G-L) & Analytic & 7.075482 & $4.838\times 10^{-16}$ & 1.892028 & 3.864327 & 1.319127 & 1/2 \\
CGC(L-G-L) & Analytic & 6.885206 & $6.103\times 10^{-16}$ & 2.519624 & 2.418858 & 1.946724 & 2/2 \\
CGC(L-G-R) & Analytic & 1.876238 & $1.724\times 10^{-16}$ & 0.290302 & 0.722734 & 0.863202 & 1/2 \\
CGC(L-G-R) & Analytic & 7.969148 & $7.343\times 10^{-16}$ & 0.917898 & 5.560451 & 1.490799 & 2/2 \\
CGC(R-G-L) & Analytic & 6.139305 & $6.949\times 10^{-16}$ & 1.354211 & 4.459709 & 0.325385 & 1/2 \\
CGC(R-G-L) & Analytic & 4.099608 & $1.285\times 10^{-15}$ & 0.247515 & 1.823477 & 2.028616 & 2/2 \\
CGC(R-G-R) & Analytic & 4.661837 & $7.073\times 10^{-16}$ & 2.562411 & 1.318116 & 0.781310 & 1/2 \\
CGC(R-G-R) & Analytic & 8.905325 & $1.187\times 10^{-15}$ & 1.455715 & 4.965069 & 2.484541 & 2/2 \\
\midrule
CCC(R-L-R) & Numeric & 1.965618 & $3.844\times 10^{-16}$ & 0.131140 & 0.674512 & 1.159966 & 1/1 \\
CGC(L-G-L) & Numeric & 6.885206 & $1.365\times 10^{-15}$ & 2.519624 & 2.418858 & 1.946724 & 1/1 \\
CGC(L-G-R) & Numeric & 1.876238 & $5.738\times 10^{-15}$ & 0.290302 & 0.722734 & 0.863202 & 1/1 \\
CGC(R-G-L) & Numeric & 4.099608 & $5.705\times 10^{-14}$ & 0.247515 & 1.823477 & 2.028616 & 1/1 \\
CGC(R-G-R) & Numeric & 4.661837 & $9.519\times 10^{-16}$ & 2.562411 & 1.318116 & 0.781310 & 1/1 \\
\bottomrule
\end{tabular}
\end{table}

Table~\ref{tab:solution_comparison} quantifies this difference for the test case.
The analytic method finds 10 feasible solution branches, while the numeric method finds only 5 solutions (one per candidate path).
This demonstrates a key advantage of the analytic approach: by solving the boundary constraint algebraically, it systematically enumerates all feasible branches without requiring exhaustive multi-start initialization.

\subsection{Computational Accuracy and Efficiency}
Table~\ref{tab:performance_comparison} summarizes the computational performance comparison between the analytic and numeric methods over 100 random test cases with $U_{\max}=2.0$.
Here, random test cases are generated by randomly sampling initial and target positions $\vct{X}_0, \vct{X}_f\in\Sph^2$ on the unit sphere, along with randomly oriented tangent (velocity direction) vectors $\vct{T}_0, \vct{T}_f$ at each position.
The corresponding Sabban frames are constructed as $R_0=[\vct{X}_0, \vct{T}_0, \vct{N}_0]$ and $R_f=[\vct{X}_f, \vct{T}_f, \vct{N}_f]$, where $\vct{N}_0=\vct{X}_0\times\vct{T}_0$ and $\vct{N}_f=\vct{X}_f\times\vct{T}_f$ complete the orthonormal frames.
The results reveal three key advantages of the analytic method: 

1) The analytic method achieves machine precision accuracy (mean error $\sim 10^{-16}$) while the numeric method exhibits iterative errors (mean error $\sim 10^{-14}$).

2) Under the same computational environment, the analytic solver is approximately 717 times faster than the numeric approach, with mean computation time of $2.24\times 10^{-3}$ seconds compared to $1.61$ seconds for the numeric solver.
This dramatic speedup arises because the analytic method solves the boundary constraint algebraically through closed-form root finding, avoiding the iterative least squares optimization loop required by the numeric approach.

3) The analytic method finds twice as many feasible solutions on average, demonstrating its completeness advantage in systematically enumerating all solution branches.

\begin{table}[htbp]
\centering
\caption{Performance comparison between analytic and numeric methods over 100 random test cases with $U_{\max}=2.0$.}
\label{tab:performance_comparison}
\begin{tabular}{lccc}
\toprule
\textbf{Metric} & \textbf{Analytic} & \textbf{Numeric} & \textbf{Ratio} \\
\midrule
Time (s) - Mean & $2.24\times 10^{-3}$ & $1.61$ & $717.22\times$ \\
Feasible Solutions - Mean & $10.10$ & $5.05$ & $2.00\times$ \\
Error (mean) & $6.387\times 10^{-16}$ & $7.269\times 10^{-14}$ & -- \\
Error (median) & $5.603\times 10^{-16}$ & $1.081\times 10^{-15}$ & -- \\
\bottomrule
\end{tabular}
\end{table}

\subsection{Parameter Sensitivity}
To assess the robustness of the analytic method across different problem regimes, we evaluate performance over a range of $U_{\max}$ values from $0.5$ to $3.0$.
Table~\ref{tab:umax_comparison} summarizes the results over 100 random test cases for each $U_{\max}$ value.

\begin{table}[htbp]
\centering
\caption{Performance comparison across different $U_{\max}$ values over 100 random test cases. The analytic method consistently achieves machine precision accuracy and finds approximately twice as many feasible solutions as the numeric method, while being orders of magnitude faster.}
\label{tab:umax_comparison}
\begin{tabular}{lcccccc}
\toprule
\textbf{$U_{\max}$} & \textbf{Method} & \textbf{Time (s)} & \textbf{Feasible} & \textbf{Error (mean)} & \textbf{Error (med)} & \textbf{Speedup} \\
\midrule
0.5 & Analytic & $0.0060$ & $6.8$ & $6.02\times 10^{-16}$ & $5.46\times 10^{-16}$ & -- \\
 & Numeric & $6.696$ & $3.4$ & $4.53\times 10^{-14}$ & $7.99\times 10^{-16}$ & $1120.3\times$ \\
\addlinespace
1.0 & Analytic & $0.0070$ & $9.0$ & $5.88\times 10^{-16}$ & $5.45\times 10^{-16}$ & -- \\
 & Numeric & $6.429$ & $4.5$ & $4.75\times 10^{-14}$ & $9.73\times 10^{-16}$ & $918.4\times$ \\
\addlinespace
1.5 & Analytic & $0.0076$ & $10.1$ & $5.17\times 10^{-16}$ & $4.47\times 10^{-16}$ & -- \\
 & Numeric & $6.074$ & $5.0$ & $5.65\times 10^{-14}$ & $1.08\times 10^{-15}$ & $797.4\times$ \\
\addlinespace
2.0 & Analytic & $0.0081$ & $10.1$ & $6.39\times 10^{-16}$ & $5.60\times 10^{-16}$ & -- \\
 & Numeric & $6.907$ & $5.0$ & $7.27\times 10^{-14}$ & $1.08\times 10^{-15}$ & $848.1\times$ \\
\addlinespace
2.5 & Analytic & $0.0068$ & $10.1$ & $5.85\times 10^{-16}$ & $5.33\times 10^{-16}$ & -- \\
 & Numeric & $5.402$ & $5.0$ & $7.06\times 10^{-14}$ & $1.37\times 10^{-15}$ & $790.9\times$ \\
\addlinespace
3.0 & Analytic & $0.0083$ & $9.8$ & $6.20\times 10^{-16}$ & $5.29\times 10^{-16}$ & -- \\
 & Numeric & $6.952$ & $4.9$ & $6.64\times 10^{-14}$ & $1.41\times 10^{-15}$ & $841.4\times$ \\
\bottomrule
\end{tabular}
\end{table}

The results demonstrate several key observations:

1) The analytic method maintains consistent machine precision accuracy across all $U_{\max}$ values, independent of the problem regime, while the numeric method exhibits errors approximately two orders of magnitude larger.

2) The analytic solver achieves substantial speedup factors ranging from approximately $790\times$ to $1120\times$ across the tested parameter range, with computation times remaining consistently below $0.01$ seconds.
The speedup is more pronounced for smaller $U_{\max}$ values, likely due to the increased complexity of the numeric optimization landscape when the turn radius becomes larger.
Note that $\Rturn=1/\sqrt{1+U_{\max}^2}$ decreases with increasing $U_{\max}$.

3) The analytic method consistently finds approximately twice as many feasible solutions as the numeric approach across all parameter values, with the number of feasible solutions increasing from $6.8$ to approximately $10$ as $U_{\max}$ increases from $0.5$ to $1.5$, then stabilizing around $10$ for higher values.
This pattern reflects the geometric constraint structure: smaller $U_{\max}$ values correspond to larger turn radii, which can limit the feasible solution space, while moderate to large $U_{\max}$ values admit richer solution sets that the analytic method successfully enumerates.

4) The computational time for the analytic method remains remarkably stable regardless of $U_{\max}$, demonstrating that the closed-form algebraic approach is insensitive to problem geometry, whereas the numeric method shows more variation in computation time depending on the optimization landscape complexity.
\section{Conclusion}
This paper proposes a unified analytic computational approach for spherical Dubins CGC and CCC paths.
The main contributions are twofold.
First, the three-dimensional Dubins path boundary value problem is formulated from a Lie group perspective, representing the vehicle configuration as a rotation matrix in $\SO(3)$ and leveraging the Sabban frame dynamics to express the boundary condition as an exponential product equation on the Lie group.
Second, based on differential geometric properties of rotation matrices, the complex three-dimensional boundary value problem is reduced to solving a simple quadratic polynomial equation.
This reduction exploits the axis-fixing property of rotations and dot product invariance to eliminate two of the three unknown segment parameters, converting the boundary constraint into a scalar trigonometric equation that admits a closed-form solution through half-angle substitution.
Numerical simulation analysis shows that the unified analytic solver achieves machine precision accuracy with errors on the order of $10^{-16}$, is approximately $717$ times faster than numerical methods under the same computational environment, and systematically enumerates all feasible solution branches without requiring exhaustive multi-start initialization.
The method provides closed-form solutions for optimal path computation in the regime where turning radius $\Rturn \in (0, 1/2]$, corresponding to $U_{\max} \geq \sqrt{3}$.

\clearpage
\section*{Appendix: Closed-form coefficients for all path types}
\label{app:coeffs}
\setcounter{equation}{0}
\renewcommand{\theequation}{A\arabic{equation}}
This appendix records the closed-form coefficients $\alpha$, $\beta$, and $\gamma$ for all CGC and CCC path types.
We use the convention $R\triangleq R_0^{\T}R_f$ for the normalized target rotation and denote its entries by $R_{ij}$.
The mapping between the letters and the saturated controls is
\begin{equation}
\label{eq:RL_controls}
R:\ u=+\Umax,\qquad L:\ u=-\Umax,\qquad G:\ u=0.
\end{equation}

\subsection{CGC path types}
For CGC path types with $(u_1,u_2,u_3)$ where $u_2=0$ and $u_1,u_3\in\{\pm\Umax\}$, the scalar feasibility equation \eqref{eq:trig_scalar} takes the form $\alpha+\beta\cos\theta_1+\gamma\sin\theta_1=0$.
The coefficients for all four CGC path types are:
\begin{equation}
\label{eq:cgc_all_coeffs}
\begin{aligned}
\text{(RG\!L)}\qquad
\alpha_{RG\!L} &= -1-\Umax\big(\Umax+\Umax R_{11}-R_{13}+R_{31}\big)+R_{33},\\
\beta_{RG\!L}  &= \Umax\big(-R_{13}+\Umax(R_{11}-\Umax R_{31}+R_{33})\big),\\
\gamma_{RG\!L} &= \Umax\sqrt{1+\Umax^2}\,(\Umax R_{21}-R_{23}),\\[0.7ex]
\text{(RG\!R)}\qquad
\alpha_{RG\!R} &= -1 + R_{33} + \Umax\!\Big(R_{13}+R_{31}+(-1+R_{11})\Umax\Big),\\
\beta_{RG\!R}  &= \Umax\!\Big(-R_{13}+\Umax\big(-R_{11}+R_{33}+R_{31}\Umax\big)\Big),\\
\gamma_{RG\!R} &= -\Umax\sqrt{1+\Umax^2}\,\Big(R_{23}+R_{21}\Umax\Big),\\[0.7ex]
\text{(LG\!R)}\qquad
\alpha_{LG\!R} &= -1 + R_{33} - \Umax\!\Big(R_{13}-R_{31}+\Umax+R_{11}\Umax\Big),\\
\beta_{LG\!R}  &= \Umax\!\Big(R_{13}+\Umax\big(R_{11}+R_{33}+R_{31}\Umax\big)\Big),\\
\gamma_{LG\!R} &= \Umax\sqrt{1+\Umax^2}\,\Big(R_{23}+R_{21}\Umax\Big),\\[0.7ex]
\text{(LG\!L)}\qquad
\alpha_{LG\!L} &= -1 + R_{33} - \Umax\!\Big(R_{13}+R_{31}+\Umax-R_{11}\Umax\Big),\\
\beta_{LG\!L}  &= \Umax\!\Big(R_{13}-\Umax\big(R_{11}-R_{33}+R_{31}\Umax\big)\Big),\\
\gamma_{LG\!L} &= \Umax\sqrt{1+\Umax^2}\,\Big(R_{23}-R_{21}\Umax\Big).
\end{aligned}
\end{equation}

\subsection{CCC path types}
For CCC path types with $(u_1,u_2,u_3)$ where $u_1,u_2,u_3\in\{\pm\Umax\}$, the scalar feasibility equation \eqref{eq:trig_scalar} takes the form $\alpha+\beta\cos\theta_1+\gamma\sin\theta_1=0$.
The coefficients for the two alternating CCC path types are:
\begin{equation}
\label{eq:ccc_all_coeffs}
\begin{aligned}
\text{(RL\!R)}\qquad
\alpha_{RL\!R} &= -\big((\Umax^2-1)\big(\Umax\big((R_{11}-1)\Umax+R_{13}+R_{31}\big)+R_{33}-1\big)\big),\\
\beta_{RL\!R}  &= 2\Umax\big(\Umax(-R_{11}+R_{31}\Umax+R_{33})-R_{13}\big),\\
\gamma_{RL\!R} &= -2\Umax\sqrt{1+\Umax^2}(R_{21}\Umax+R_{23}),\\[0.7ex]
\text{(LR\!L)}\qquad
\alpha_{LR\!L} &= -\big((\Umax^2-1)\big(-\Umax(-R_{11}\Umax+R_{13}+R_{31}+\Umax)+R_{33}-1\big)\big),\\
\beta_{LR\!L}  &= 2\Umax\big(R_{13}-\Umax(R_{11}+R_{31}\Umax-R_{33})\big),\\
\gamma_{LR\!L} &= 2\Umax\sqrt{1+\Umax^2}(R_{23}-R_{21}\Umax).
\end{aligned}
\end{equation}

\section*{Acknowledgments}
This work was supported by the National Natural Science Foundation of China under Grant No. 12472359 and No. U2441205.
\bibliography{sample}

\end{document}